\documentstyle[preprint,aps,epsfig]{revtex}

\begin{document}
\title{
Quantum wires and dots driven by intense surface acoustic waves and the quantum
attenuation of sound in an electron plasma }

\author{A.~O.~Govorov$^*$, A.~V.~Kalameitsev$^{**}$, and V.~M.~Kovalev}
\address{Institute of Semiconductor Physics,
Russian Academy of Sciences, \\
Siberian Branch, 630090 Novosibirsk, Russia}
\maketitle
\begin{abstract}
We develop a quantum theory of the nonlinear interaction between
intense surface acoustic waves and electrons of a quantum well in
the regime of moving quantum wires and dots. The quantum nonlinear
interaction qualitatively differs from the classical one. In a
system of electron wires driven by an acoustic wave the sound
attenuation strongly decreases with increasing the sound
intensity. However, even in the quantum limit the sound
attenuation in the electron system remains nonzero due to residual
scattering by impurities in the lowest subband. In the case of
electron dots formed by two waves moving in the perpendicular
directions, the calculated dissipation vanishes at high sound
intensity because of the fully quantized spectrum. Besides, moving
quantum dots can be formed by a single acoustic wave propagating
along a system of etched quantum wires. We show that the quantum
regime of the attenuation can exist in high-mobility GaAs quantum
wells where the heating can not destroy the quantum state.
\end{abstract}
\pacs{PACS numbers: xxxx}

\section{Introduction}
\label{intro}

Quantum wires, channels, and dots are very intriguing objects in the modern
solid-state physics. They display new physical properties due to reduced
dimensionality. The electric conductance of narrow channels in a semiconductor
nanostructures demonstrates the quantization \cite{Wees-WharamJPCM} described by
the Landauer-B\"uttiker theory \cite{Landauer-Buttiker}.
More recently the quantized conductance has also been found for the thermal
transport in free-standing quantum wires \cite{thermal}.
Furthermore the quantum wires and dots are attracting significant theoretical
interest due to the enhanced many-particle interactions
\cite{theory,Govorov91,Hawrylak}.

Experimentally, semiconductor quantum wires and channels are
realized or by the electrostatic method or in etched
nanostructures \cite{Wees-WharamJPCM,wires}. Recently, another
method to create electron wires has been demonstrated
\cite{Rotter,Govorov00}. This method involves an intense surface
acoustic wave (SAW) in a hybrid structure  \cite{Rotter}
containing a semiconductor quantum-well and a strongly
piezoelectric (PE) host crystal (Fig.~1a). In the experiments
\cite{Rotter} a formerly homogeneous two-dimensional (2D) electron
gas in the hybrid structure turns into moving electron wires due
to a very strong piezoelectric potential induced by a SAW. This
effect has been demonstrated in the above-mentioned experiments
\cite{Rotter} for the classical regime of the acoustoelectric (AE)
interaction at room temperature. In the room-temperature electron
system the dissipation of the SAW energy increases with increasing
the sound intensity and approaches its maximum in the limit of the
high amplitude of a SAW \cite{Rotter,Govorov00,3Dnonlin}.

For low temperatures and a sufficiently strong PE potential of a
SAW, the electron spectrum of wires can be quantized in the
direction of the SAW momentum (Fig.~1b)
\cite{Keldysh,Popov,Laikhtman}. For the first time the electron
quantization effects in an acoustic wave have been discussed by
Keldysh \cite{Keldysh}. Our estimations show that the lateral
quantization could be observed in low-temperature experiments on
the recently fabricated hybrid structures.

Here we develop a quantum theory of the AE interaction in a 2D
system in the strongly nonlinear regime when intense SAW's split a
2D electron system into moving quantum wires and dots. We model
the AE interaction from first principles for the case of electron
scattering by impurities and use time-dependent perturbation
theory. In the quantum regime the SAW absorption due to electrons
reflects the density of states in quantum wires. When a few
quantum subbands are occupied the SAW energy dissipation comes
mostly from the inter-subband transitions assisted by impurities.
Furthermore, even in the strictly 1D case the SAW dissipation in
quantum wires remains finite. The residual absorption of SAW in
the 1D wires originates from the intra-subband transitions in the
direction perpendicular to the SAW momentum. Our numerical results
show that the residual SAW absorption in high-quality GaAs quantum
wells is small enough and, so, the heating effect is weak in the
quantum limit. In the case of two SAW's moving in the
perpendicular directions, an electron system can be split into
moving dots. In contrast to the case of wires, the SAW dissipation
due to electron dots decreases and approaches zero in the quantum
limit.

Many experimental papers report on interactions of SAW's with the
quantum Hall effect systems and lateral nanostructures
\cite{Wixforth89,Talyanskii}. However, most of those studies
relate to the linear regime of the AE interaction. A linear theory
of the AE interaction in a 2D system and various nanostructures
was developed in Refs.~\cite{2Dlin,nanolin}. Also, theoretical
aspects of single-electron transport through a quantum-point
contact were discussed in Ref.~\cite{Aizin}.

The nonlinear AE interaction in the classical regime in bulk
crystals was widely discussed theoretically \cite{3Dnonlin}.
Recently a theory of the nonlinear AE effects has also been
considered for the case of a classical 2D plasma in hybrid
structures \cite{Govorov00}. For the case of 3D plasma, the
attenuation of sound in the presence of quantization in one
direction was studied by Laikhtman and Pogorel'skii
(Ref.~\cite{Laikhtman}). They used the quasi-classical approach
based on the Boltzmann equation and have predicted the quantum
oscillations of the attenuation due to the density of states. Here
we study the nonlinear quantum AE interactions for the case of a
2D system in the regimes of moving quantum wires and dots.
Furthermore, we develop a quantitative theory based on the
density-matrix motion equation. The quantum-mechanics treatment of
this problem is  necessary for the quantitative theory, which
takes into account non-diagonal elements of the density matrix and
the quantum dynamical screening of the impurity potential. In
particular, the screening effect is important for the model of
background Coulomb impurities. The calculations performed in the
Coulomb-impurity model without the screening effect lead to the
infinitely large sound attenuation. Basing on our numerical data
we can conclude that the quantum attenuation regime is possible in
high-mobility GaAs structures despite heating. It is in contrast
to the AE interaction in the 3D plasma where the heating effect is
predicted to be strong (Ref.~\cite{Laikhtman}). In addition, we
find that the sound dissipation in the case of driven quantum dots
behaves differently compared to the nonlinear regimes in the 3D
plasma and to the quantum wires.

\section{ General formalism }
\label{model}

The interaction between a SAW and a 2D electron system is characterized by the
SAW energy absorbed in an electron system per unit time and aria

\begin{equation}
Q=<{\bf j}_s({\bf r},t){\bf E}_{SAW}(x,t)>_{\bf r},
\label{Q1}
\end{equation}
where $<...>_{\bf r}$ means averaging over surface aria of a macroscopic sample.
${\bf r}=(x,y)$ is the in-plane coordinate and $t$ is the time.
The 2D electron current ${\bf j}_s$ in the plane of a quantum well is induced by
the PE field of a SAW, ${\bf E}_{SAW}(x-c_st)$. The Rayleigh SAW propagates in
the $x$-direction with a
velocity $c_s$ and interacts with a 2D plasma (Fig.~1).
In this case, the in-plain component of ${\bf E}_{SAW}$ is parallel to the $x$-
axis.
Eq.~\ref{Q1} can be used for a description of the SAW dissipation in the
nonlinear regime in the limit of a weak AE coupling $K_{eff}^2\ll1$
\cite{Govorov00}.

To find the electron current in the presence of the SAW piezoelectric field, we
use the equation of motion for the one-particle density matrix $\hat{f}$ within
the framework of the self-consistent field approximation \cite{QuinnJJ},

\begin{equation}
\frac{\partial \hat{f}}{\partial t} +
\frac{i}{\hbar}[\hat{H},\hat{f}]=(\frac{\partial \hat{f}}{\partial
t})_{Collisions},
\label{MD1}
\end{equation}
where $\hat{H}$ is the electron Hamiltonian.

The conventional approach to the equation (2) is based on the constant
relaxation time $\tau$ and the assumption that the collision term in
Eq.~\ref{MD1} is proportional to the divination from the local thermal
equilibrium \cite{QuinnJJ}.
However, this approach is not effective in our case.
By integrating Eq.~\ref{MD1} and the similar equation for the current we can
show that the SAW-energy dissipation in the regime of well separated wires is
$Q=<j_sE_{SAW}>_{\bf r}=m^*c_s^2N_s/\tau$, where $m^*$ is the electron mass.
Here $N_{s}=<n_s(x,t)>_x$ is the averaged 2D density and $n_s({\bf r},t)$ is the
local density. This result coincides with the Weinreich relation
\cite{Weinreich} obtained within the Drude model \cite{Govorov00,3Dnonlin}.
Thus, the absorption $Q$ found in the simplified model does not exhibit any
physical features of the lateral quantization.

To describe the AE interaction in the quantum wire regime we need to model
scattering from the first principles. For low temperatures the main contribution
to $Q$ comes from the electron scattering by impurities \cite{Laikhtman,SNOSKA}.
So, we introduce the potential of randomly located impurities directly into the
Hamiltonian in the right-hand side of Eq.~\ref{MD1}.
The Hamiltonian describing the in-plane single electron motion in the presence
of SAW piezoelectric potential $\Phi_{SAW}$ is

\begin{equation}
\hat{H}=\frac{\hat{{\bf p}}^{2}}{2m} + e\Phi_{SAW}(x-c_{s}t) +
U_{ind}({\bf r},t) + U_{imp}({\bf r}),
\label{H}
\end{equation}
where $\hat{{\bf p}}$ is the in-plane momentum and
$e$ is the electron charge.
$\Phi_{SAW}(x-c_{s}t)=\Phi_{SAW}^0 \cos{(kx-\omega t)}$ is the piezoelectric
potential and $\omega=c_sk$.
$U_{ind}$ is the potential induced by the electrons.
$U_{imp}({\bf R})=\sum_{i}U_{0}({\bf R}-{\bf R}_{i})$, where ${\bf R}=({\bf
r},z)$ is the 3D electron coordinate, and $U_{0}({\bf R}-{\bf R}_{i})$ is the
potential of the $i$-impurity.
Here the electron motion is 2D in the plane $z=0$ whereas the impurity position
$R_i$ is a 3D vector. In a quantum well the electron motion in the $z$-direction
is described by a localized wave function $\chi_0(z)$. For simplicity, we will
neglect the width of the function $\chi_0(z)$ and consider the potential
$U_{imp}({\bf R})$ only in the plane $z=0$. Also, we assume that electrons
occupy only the ground 2D subband.
To describe the SAW dissipation we intend to use the model of background Coulomb
impurities with homogenous spatial distribution. In the following we will denote
the
average 3D density of impurities by $N_t$.

\section{Perturbation theory}
\label{PT}

First we qualitatively analyze perturbation theory for the regime
of wide wires where the quantization is negligible. The impurity
potential is assumed to be weak perturbation.  If the SAW
intensity is sufficiently low a SAW slightly modulates a 2D
plasma. In this linear regime \cite{2Dlin}, the sound absorption
coefficient $\Gamma_0=Q/I_{SAW}\propto\sigma_0\propto\tau_{tr}$
when $\sigma_0\rightarrow 0$. Here $I_{SAW}$ is the SAW intensity.
In the limit $U_0\rightarrow 0$,  $\tau_{tr}\rightarrow\infty$ and
$Q\rightarrow\infty $. It shows that the simplest first-order
perturbation theory is not valid. Indeed, it is known that the
conductivity can be calculated as an infinite series of "ladder"
diagrams \cite{Abrikosov}. The perturbation theory of
Ref.~\cite{Abrikosov} is based on the parameter $k_fl_e\gg1$,
where $k_f$ and $l_e$ are the Fermi wave vector and the electron
mean free path, respectively. We see that the first diagram in
terms of the impurity potential is not sufficient in the linear
regime of the AE interaction. The situation in the nonlinear case
is qualitatively different. For a high SAW intensity a 2D plasma
forms well separated wires. When $I_{SAW}\rightarrow\infty$, the
SAW dissipation due to classical wires becomes saturated and equal
to $Q_{max}= m^*c_s^2*N_s/\tau_{tr}$
\cite{Govorov00,3Dnonlin,Qmax}. One can see that
$Q_{max}\propto1/\tau_{tr}\rightarrow 0$ when $U_0 \rightarrow 0$.
It means that we can calculate $Q_{max}$ by using the first
diagram in terms of the parameter $U_0^2$. In other words, in the
case of wires the SAW potential confines strongly electrons and we
can take into account the impurity scattering using simplest
first-order perturbation theory. In the following we will use this
fact to calculate the SAW absorption.

In the regime of wires, it is convenient to introduce the new moving coordinate
$x'=x-c_st$. In the moving coordinate system ${\bf r}'=(x',y)$, the strong SAW
potential $e\Phi_{SAW}(x')$ is time-independent but the impurity potential
$U_{imp}({\bf r}',t)$ depends on the time.
By changing ${\bf r}\rightarrow{\bf r}'$ we change
$\hat{f}$ to $\hat{f}'$, where $\hat{f}'$  is the density matrix in the moving
system of coordinates. Obviously, the motion equation (\ref{MD1}) is conserved
when ${\bf r}\rightarrow{\bf r}'$.

The quantum wire width $l_0$ is typically much less than the
acoustic-wave length $\lambda\sim1~\mu m$ and, thus, the
interaction between the quantum wires can be neglected.  So, for
convenience we will consider the density matrix for a single wire.
In the moving coordinate system the electron current of a single
wire contains only the random term $\delta{\bf j}_{w}$ which is
induced by impurities. The dissipation is given by

\begin{equation}
Q=\frac{1}{\lambda} \int d x'
<\delta {\bf j}_{w}({\bf r'},t){\bf E}_{imp}({\bf r}',t)>_{{\bf R}_{i}},
\label{Q2}
\end{equation}
where
${\bf E}_{imp}=-\nabla_{{\bf r}'}U_{imp}({\bf r}',t)/e$ is the random electric
field of impurities.
$<..>_{{\bf R}_{i}}$ means averaging over random positions of impurities.
Eqs.~\ref{Q1} and \ref{Q2} are equivalent because the dissipation is independent
of the coordinate system. The electron density and current in the static system
can be written as $n_s({\bf r},t)=n_{0}(x-c_st)+\delta n_{1}({\bf r},t)$ and
${\bf j}_s({\bf r},t)={\bf j}_0(x-c_st)+\delta {\bf j}_1({\bf r},t)$,
respectively.
Here
$\delta n_{1}$ and $\delta {\bf j}_1$ are random terms. The regular
contributions are connected by ${\bf j}_0=e{\bf c}_sn_{0}(x-c_st)$.
The dissipation given by Eq.~\ref{Q1} is $Q=\int{\bf j}_0{\bf E}_{SAW}dx$
because
$<\delta{\bf j}_1({\bf r},t){\bf E}_{SAW}>_{{\bf R}_i}=0$.
We can directly show that
$\int{\bf j}_0{\bf E}_{SAW}dx=
\int <\delta{\bf j}_{w}{\bf E}_{imp}>_{{\bf R}_i}dx'$ if we multiply the
equation of motion (\ref{MD1}) by the current operator and then take the trace.
Here we note that the electron density $n_s$ and the currents relate to a single
wire.

On the next step we consider linear response of a wide electron
wire to the potential of "moving" impurities $U_{imp}({\bf
r}',t)$. In wide wires we can neglect the lateral quantization.
In a small part of the electron wire, the electron plasma can be
regarded as a homogeneous 2D Fermi gas.  The linear response of a
homogeneous 2D plasma to the potential $U_{imp}({\bf r}',t)$ can
be calculated from Eq.~\ref{MD1} \cite{QuinnJJ}.  The dissipation
in the a small part of the channel near the point $x_0'$ is
$\delta Q=\int_{x_0'}^{x_0'+\delta x'} <\delta{\bf j}_{w} {\bf
E}_{imp}>_{{\bf R}_i}dx'$. Then, using the dynamic 2D electron
conductivity $\sigma_s({\bf q},\omega)$, we obtain $\delta
Q=\delta x'n_s(x')c_s^2m^*/\tau_{tr}$. After integration over all
parts of the wire we again come to the above-mentioned equation
for $Q_{max}$.

In the absence of impurities, the electron-electron scattering in the moving
coordinate system results in thermal equilibrium. In this case the density
matrix is equal to the Fermi distribution function $\hat{f}^{0}$ which depends
on the electron temperature $T_e$. In the presence of weak impurity scattering
the density matrix becomes equal to $\hat{f}'=\hat{f}^0+\delta\hat{f}'$.
The spectrum of quantum wires has the 1D subband structure:
$E_{p_y,\alpha}=E_{\alpha}+p_y^2/2m^*$, where $p_y$ and
$\alpha$ are the $y$-component of the momentum and the subband index;
$\alpha=0,1,2,...$ . In the self-consistent field approximation, the electrons
in an ideal wire move in the potential $e\Phi_{SAW}(x')+V_{ind}(x')$, where
$V_{ind}(x')$ is the term induced by the Coulomb repulsion. The single-electron
wave functions will be denoted as $\psi_{\alpha}(x') e^{ip_yy}$. Also, it is
convenient to use the parabolic approximation for the SAW potential,
$e\Phi_{SAW}=e\Phi_{SAW}^0+m^*\Omega_0^2{x'}^2/2$, which is valid near the
center of a wire.  Here $\Omega_0=k\sqrt{|e|\Phi_{SAW}^0}/m^*$.

In the coordinate system $({\bf r}',t)$, the potential $U_{imp}({\bf r}',t)$ at
$z=0$ plays the role of perturbation.  By using a linear response theory
\cite{QuinnJJ} we find the Fourier transform of the charge density perturbation
induced by the time-dependent impurity potential

\begin{eqnarray}
\delta\rho(x',q_{y},\omega)=
e\sum_{\alpha,\beta}\psi_{\alpha}(x')
\psi_{\beta}(x')
<\psi_{\alpha}(x')|\tilde{U}_{imp}(x',q_{y},\omega)|\psi_{\beta}(x')>
\Pi_{\alpha,\beta}(\omega,q_{y}),
\label{rho}
\\
\Pi_{\alpha,\beta}(\omega,q_{y})=\int\frac{d
p_y}{\pi\hbar}\frac{f^{0}_{p_y,\alpha}-f^{0}_{p+q_{y},
\beta}}{\hbar\omega+E^{0}_{p_y,\alpha}-E^{0}_{p_y+q_{y},\beta}+i0}
\nonumber
\end{eqnarray}
where  $q_y$ and $\omega$ appear after Fourier transformation in terms of $y$
and $t$, respectively.  Here we have neglected weak electron-electron
scattering.  $\tilde{U}_{imp}$ is the screened impurity potential at $z=0$.

Using Eqs.~\ref{Q2} and \ref{rho} we obtain

\begin{eqnarray}
\label{Q3}
Q=\frac{1}{\lambda}\frac{N_{t}}{(2\pi)^2\hbar}\sum_{\alpha,\beta}
\int dq_xdq_y D_{\alpha,\beta}(q_{y},c_sq_{x})
<|U_{0}(q_{x},q_{y},Z_i)|^{2}>_{Z_i}*
\\
|A_{\alpha,\beta}(q_{x})|^{2}
\int dp_y c_s q_{x}(f^{0}_{p_y,\alpha}-f^{0}_{p_y+q_{y},\beta})
\delta(\hbar c_sq_{x}+E^{0}_{p_y,\alpha}-E^{0}_{p_y+q_{y},\beta}),
\nonumber
\end{eqnarray}
where $N_t$ is the 3D impurity density.
$A_{\alpha,\beta}(q_{x})=\int dx
\psi_{\alpha}(x)\psi_{\beta}(x)e^{iq_{x}x}$ and
$D_{\alpha,\beta}(q_{x},q_{y},\omega)$ is the screening factor.
The single-impurity potential is given by $U(q)=2\pi e^2/(\epsilon
|q|)e^{-|Z_i||q|}$; $<|U_{0}|^{2}>_{Z_i}=(2\pi)^2
e^4/(\epsilon^2|q|^3)$ and ${\bf q}=(q_x,q_y)$. The screening
factor for the impurity potential in a quantum wire with many
occupied subbands should be found from the system of linear
equations \cite{QuinnJJ}. In the simplest case of the lowest
occupied 1D subband we obtain
$D_{0,0}(q_{y},\omega)=1/|\epsilon_{0,0}(q_y,\omega)|^2$, where
$\epsilon_{0,0}(q_y,\omega)=1+V_{0,0}(q_y)\Pi_{0,0}(q_y,\omega)$
and $V_{0,0}$ is the matrix element of the Coulomb potential.

Equation \ref{Q3} describes inter- and intra-subband transitions
in a quantum wire induced by "moving" impurities. The momentum and
energy transfers in these transitions are ${\bf q}=(q_x,q_y)$ and
$\hbar{\bf qc_s}=\hbar c_sq_x$. In Fig.~1c we show electron
transitions which contribute to $Q$. The difference between
Eq.\ref{Q3} and the quasi-classical results of
Ref.\cite{Laikhtman} is the screening factor $D_{\alpha,\beta}$
calculated from the quantum equations. In the case of a few
occupied subbands, the screening factor has a complex structure.

Electron scattering by impurities also leads to the spatial shift
of the electron distribution from the center of a quantum wire.
This shift directly relates to the dissipation. Multiplying the
equation of motion (\ref{MD1}) by the current operator and then
taking the trace, we can obtain the steady-state condition
$Q/c_s=e(1/\lambda)\int<n_sE_{SAW}>_{{\bf R}_i}dx'$, where $n_s$
relates to a single wire. Here the term $e\int <n_sE_{SAW}>_{{\bf
R}_i}dx'$  is the averaged force acting on the electrons in the
wire and $\lambda Q/c_s$ plays the role the friction force. The
nonzero quantity $e\int<n_sE_{SAW}>_{{\bf
R}_i}dx'=-m^*\Omega_0^2\int <x'n_s>_{{\bf R}_i}dx'$, i.e. it is
proportional to the shift of the electron distribution from the
wire center. We have directly calculated the shift
$e\int<n_sE_{SAW}>_{{\bf R}_i}dx'$ using the second-order
perturbation theory for the density matrix $\hat{f}'$.  The
obtained formula was consistent with the steady-state condition
$Q/c_s=e(1/\lambda)\int<n_sE_{SAW}>_{{\bf R}_i}dx'$ and
Eq.~\ref{Q3} for $Q$.

\section{Numerical results for the quantum wire regime}
\label{NR}

In Figs.~2 and 3 we show the calculated SAW absorption as a function of the
potential amplitude $\Phi_{SAW}^0$ for various temperatures. We choose the
following parameters $\lambda=1~\mu m$,
$c_s=3.9*10^5~cm/s$, $N_L=4*10^{5}~cm^{-1}$,
$N_s=N_L/\lambda=4*10^9~cm^{-2}$, $N_t=4.1*10^{14}~cm^{-3}$, $m^*=0.07 m_0$, and
$\epsilon=12.5$.
The Fermi energy in the 1D limit is $E_{f}^{1D}\simeq2~meV$.
The parameter $N_t$ is found from the low-temperature mobility of a 2D
homogeneous gas $\mu_{2D}$.  For the mobility we take $\mu_{2D}=3*10^6~cm^2/Vs$
at the 2D density $3*10^{11}~cm^{-2}$.
$\mu_{2D}$ was calculated by using the Born approximation and involving the
screened impurity potential \cite{Davies}.

Figure 2 shows the SAW absorption for the case of electron
scattering by background impurities with a short-range potential.
The 3D density and the potential of $\delta$-impurities correspond
to the 2D mobility $3*10^6~cm^2/Vs$. To calculate $\mu_{2D}$ we
take into account a finite width of a quantum well in the
$z$-direction. In the case of $\delta$-impurities we use the
equation similar to Eq.~\ref{Q3} and involve a few lowest 1D
subbands. To demonstrate a general behavior of $Q(\Phi_0^{SAW})$,
we use the single-particle wave functions, neglecting
self-consistent Coulomb effects. The function $Q(\Phi_{SAW}^0)$
reflects the density of states in a wire. For relatively small SAW
potentials electrons occupy a few 1D subbands and the main
contribution to $Q$ comes from inter-subband transitions induced
by impurities (Fig.~1c). The contribution of inter-subband
transitions is responsible for the absorption in the
quasi-classical and classical regimes. At
$\Phi_{SAW}^0\simeq0.08~V$ the Fermi level coincides with the
bottom of the first exited subband $\alpha=1$ which leads to a
maximum in the function $Q(\Phi_{SAW}^0)$. When
$\Phi_{SAW}^0\rightarrow0$, $Q\rightarrow Q_{max}$. For the small
$\Phi_{SAW}^0$, the features related to higher subbands are not
seen because of finite temperature. For the high $\Phi_{SAW}^0$
the electrons occupy only the ground subband. In the case of
$\delta$-impurities the dissipation $Q$ is an increasing function
of the potential amplitude in the limit of
$\Phi_{SAW}^0\rightarrow\infty$. In the latter case the SAW
absorption occurs only due to intra-subband electron transitions
(Fig.~1c).

Now we calculate the function $Q(\Phi_{SAW}^0)$ in the model of
dynamically screened Coulomb impurities. We use the
self-consistent field approximation and the variational method to
find the wave functions $\psi_\alpha(x')$. In the Coulomb-impurity
model, the SAW absorption strongly decreases with increasing the
SAW potential amplitude $\Phi_{SAW}^0$ (Fig.~3). At high
$\Phi_{SAW}^0$, the absorption $Q$ becomes saturated. The residual
absorption at $\Phi_{SAW}^0\rightarrow\infty$ originates from
scattering within the ground subband (Fig.~1c) and depends on the
quality of a quantum well and on the electron density. Again at
$\Phi_{SAW}^0\simeq0.2~V$ we can see the characteristic maximum in
the function $Q(\Phi_{SAW}^0)$ which reflects the electron
occupation of the second subband. The maximum of $Q(\Phi_{SAW}^0)$
in Fig.~3 is shifted to higher potentials compared with the data
of Fig.~2. The reason is the Coulomb interaction in a wire.
Inter-subband transitions for small potentials $\Phi_{SAW}^0$
result in the strong increase of $Q$.

The dynamic screening of Coulomb impurities plays the very
important role. The dissipation $Q$ calculated by Eq.~\ref{Q3} is
non-convergent in the absence of the screening effect. Also, we
find that the screening effect for intra-subband transitions is
much stronger than that for inter-subband transitions. It leads to
a strong increase of the absorption as the sound intensity
decreases. Another effect of inter-particle interaction is that
the energy spacing between 1D subbands becomes essentially less
due to Coulomb repulsion in a wire \cite{wires}. The energy
quantization in a wire is reduced by a factor of $2$ after the
inclusion of the self-consistent Coulomb effects. This effect is
seen from the data of Fig.~2 and Fig.~3.

It is interesting to compare the quantum attenuation with that in the classical
limit. At room temperature the absorption is increasing with $\Phi_{SAW}$ and
becomes saturated for high sound intensities (insert of Fig.~3)
\cite{Govorov00,3Dnonlin}. The maximal absorption depends directly on the
mobility:
$Q_{max}=m^*c_s^2*N_s/\tau_{tr}$. For the typical room-temperature parameters
$\mu=5000~cm^2/Vs$ and $N_s=4*10^9~cm^{-2}$
\cite{Rotter}, we obtain $Q_{max}\simeq2*10^5~erg/s$ which is a few orders of
magnitude larger than the calculated quantum-limit absorption.

Here we have calculated the SAW absorption assuming that the
electronic temperature in a wire is independent of $\Phi_{SAW}^0$.
However, the electron temperature $T_e$ can depend on
$\Phi_{SAW}^0$ and differ from the lattice temperature $T_{l}$.
The important factor is the phonon relaxation efficiency, which
depends on a material. The heating effect can be analyzed by the
energy balance equation $Q=P_{ph}$, where $P_{ph}$ describes the
efficiency of the electron energy relaxation due to the emission
of acoustic phonons. Involving the deformation- potential and
piezoelectric mechanisms we calculate the energy-loss rate
$P_{ph}(T_{l},T_e)$ for the driven electron wires in the
$GaAs$-based quantum well \cite{Bockelmann}. Figs.~4 and 5 show
the phonon energy-loss rate $P_{ph}$ and the calculated heating
temperature $\Delta T=T_e-T_{l}$ in the 1D limit for
$\Phi_{SAW}^0=2~V$. Here we find the electron temperature from the
equation $Q(T_e)=P_{ph}(T_{l},T_e)$ for the case of Coulomb
impurities (Fig.~5). One can see that the heating effect in the 1D
quantum regime is small in high mobility systems (Fig.~5). At the
same time, we find that at smaller SAW intensity, when the
function $Q$ rapidly increases, the heating effect in $GaAs$-wires
is stronger and can destroy the quantum state. In particular, the
maximum of the function $Q(\Phi_{SAW}^0)$ for
$\Phi_{SAW}^0\simeq0.25~V$ can vanish. For
$\Phi_{SAW}^0\simeq0.25~V$ and $Q\simeq10^4~erg/(cm^2s)$, the
roughly estimated electron temperature in wires can be about
$30~K$.

As an example, we now discuss the solution of the balance equation
$Q(T_e)=P_{ph}(T_{l},T_e)$ for $T_{l}=2~K$ (Fig.~5). One can see
that the crossing between the curves for $Q$ and $P_{ph}$ in
Fig.~4 exists when $\mu>0.8*10^6~cm^2/Vs$ (Fig.~5). So, for given
lattice temperature the solution of $Q(T_e)=P_{ph}(T_{l},T_e)$ is
not longer existing if the 2D mobility is low.   We can speculate
that, if $\mu<0.8*10^6~cm^2/Vs$, the SAW can heat the electron
wire up to the temperature about $30~K$. At $T_e\sim30~K$ the
$LO$-phonon scattering starts to play the role \cite{Bockelmann}.
Thus, the quantum regime can exist only in high-quality quantum
wells. In addition, we note that the solutions of the balance
equation shown in Figs.~4 and 5 are stable.

The predicted quantum effects can occur if the potential amplitude induced
by a SAW is sufficiently high. For $N_L=4*10^5~cm^{-1}$, $T_e=2~K$, and
$\lambda=1~\mu m$, the quantum 1D regime occurs when $\Phi_{SAW}^0>1~V$.
The amplitudes $\Phi_{SAW}^0$ up to $2~V$ have been achieved in the experiments
\cite{Rotter} on the hybrid structures with a closely-located top metallic gate
for SAW's with $\lambda=10~\mu m$. Additional enhancement of quantum effects can
be expected in hybrid structures involving even stronger piezoelectric materials
(Potassium Niobate, $K_{eff}^2\sim0.5$ \cite{potassium}) and semiconductors with
smaller effective masses ($InAs$, $m^*=0.03 m_0$).

In most of experiments on SAW's \cite{Rotter,Wixforth89} the measured quantity
is  the absorption coefficient $\Gamma=Q/I_{SAW}$, where $I_{SAW}$ is the SAW
intensity.
The calculated SAW absorption becomes saturated at high sound intensity and thus
$\Gamma$ behaves as $1/I_{SAW}$ in the limit $I_{SAW}\rightarrow\infty$
\cite{Laikhtman,3Dnonlin}.
To observe the quantum effects we suggest to measure directly the
SAW absorption because it is not decreasing in the limit
$I_{SAW}\rightarrow\infty$.  In the experiments on hybrid structures the total
measured absorption in a sample was about or more then $0.5~erg/s$
\cite{Rotter}.
In $GaAlAs$ systems, the observed absorption can be even less \cite{Wixforth89}.
The total sound absorption in a sample with a surface aria $S$ is given by
$W=S*Q$. Using the data of Fig.~3 we obtain $W\sim erg/s$ for $T=2~K$,
$\Phi_{SAW}^0>1~V$, and $S=0.1*0.1~cm^2$. Hence, the residual attenuation in the
quantum 1D limit could be experimentally observed. In experiments, the electron
plasma can be induced by the metal gate (Fig.~1) \cite{Rotter} or photogenerated
\cite{Rocke}.
Recently, the nonlinear AE interactions between SAW's and a photogenerated
plasma were observed in $GaAs$-based structures \cite{Rocke}.

\section{Quantum dots driven by acoustic waves}
\label{Dots}

In the presence of two SAW's with perpendicular wave vectors the
electron motion can be quantized in two directions. Assume that
SAW1 and SAW2 propagate in the $x$- and $y$-directions,
respectively. Their potentials are
$\Phi_{SAW1}=\Phi_{SAW1}^0\cos{(kx-\omega t)}$ and
$\Phi_{SAW2}=\Phi_{SAW2}^0\cos{(ky-\omega t)}$. The total
potential near its minimums can be approximated as
$e\Phi_{SAW1}+e\Phi_{SAW2}\simeq
e\Phi_{SAW1}^0+e\Phi_{SAW2}^0+(m^*/2)(\Omega_x^2x'^2+\Omega_y^2y'^2)$,
where the frequencies
$\Omega_{x,y}=k\sqrt{|e|\Phi_{SAW1(2)}^0/m^*}$ depend on the
intensities of the SAW's. In the moving coordinate system
$(x',y')=(x-c_st,y-c_st)$, the electron spectrum is discrete and
given by $E_{n,m}=\hbar\Omega_x(n+1/2)+\hbar\Omega_y(m+1/2)$. The
corresponding wave function will be denoted as
$\Psi_{n,m}(x',y')=\psi^x_{n}(x')\psi^y_m(y')$. In the following
we will consider a single dot and, for simplicity, neglect the
Coulomb interaction.

The electron current in a dot in the moving coordinate system can be calculated
by using Eq.~\ref{MD1}.
The $x'$-component of the random electron current induced by impurities in the
moving coordinate system is

\begin{eqnarray}
\label{j1}
\delta j_{ind,x'}(x',y',\omega)=
-\frac{i\hbar e}{2 m^*}\sum_{n,m,n_1,m_1}
[\Psi_{n,m}(x',y')\frac{\partial\Psi_{n_1,m_1}(x',y')}{\partial x'}-
\Psi_{n_1,m_1}(x',y')\frac{\partial\Psi_{n,m}(x',y')}{\partial x'}]
\\ \nonumber
<\Psi_{n_1,m_1}|U_{imp}(x',y',\omega)|\Psi_{n,m}>
\frac{f^{0}_{n,m}-f^{0}_{n_1,m_1}}{\hbar\omega+E_{n,m}-E_{n_1,m_1}+i0}.
\end{eqnarray}

The absorption of two SAW's by a single dot is given by

\begin{eqnarray}
\label{Qdot}
Q=\frac{1}{\lambda^2}\int dx'dy'<\delta{\bf j}_{ind}*{\bf E}_{imp}>_{{\bf R}_i}.
\end{eqnarray}

By using Eqs.~\ref{j1} and \ref{Qdot} we obtain

\begin{eqnarray}
\label{Qdot1}
Q=\frac{1}{\lambda^2}
\frac{1}{4\pi}N_t2\sum_{n,m,n_1,m_1}
\int dq_xdq_y <|U_0(q)|^2>_{Z_i}| A_{n,m,n_1,m_1}({\bf q})|^2
\\ \nonumber
(c_sq_x+c_sq_y)
(f^{0}_{n,m}-f^{0}_{n_1,m_1})
\delta(\hbar c_sq_x+\hbar c_sq_y+E_{n,m}-E_{n_1,m_1}),
\end{eqnarray}
where ${\bf q}=(q_x,q_y)$ and
$A_{n,m,n_1,m_1}({\bf q})=\int\Psi_{n,m}\Psi_{n_1,m_1}e^{i{\bf qr}}dxdy$.
The factor $2$ in Eq.~\ref{Qdot1} is due to the spin degeneracy.

For the case $\Phi_{SAW1}^0=\Phi_{SAW2}^0=\Phi_{SAW}^0$, the wave function
$\Psi_{n,m}\propto e^{-r^2/2l_0}$, where
$l_0=\sqrt{\hbar/\Omega_0m^*}$.
In the limit $\Phi_{SAW}^0\rightarrow\infty$ and for low temperatures, the
leading terms in Eq.~\ref{Qdot1} have the indexes $n-n_1=\pm1$ and $m=m_1$ or
$m-m_1=\pm1$ and $n=n_1$.
The asymptotic behavior of $Q$ is

\begin{eqnarray}
\label{Qdot22}
Q\propto e^{-\frac{\Omega_0^2 l_0^2}{2c_s^2}}=
e^{-\frac{\hbar\Omega_0}{2m^*c_s^2}}=
e^{-\sqrt{\frac{\Phi_{SAW}^0}{\Phi_s}}},
\end{eqnarray}
where $\Phi_s=4(m^*c_s^2)^2*m^*/[e\hbar^2k^2]$.
We see that the absorption in the limit of the large SAW intensity decreases
exponentially.

The absorption $Q$ is a sum of two contributions $Q_1$ and $Q_2$
which relate to the SAW1 and the SAW2, respectively. In the
general case $\Phi_{SAW1}^0\neq\Phi_{SAW2}^0$ and both $Q_1$ and
$Q_2$ depend on $\Phi_{SAW1}^0$ and $\Phi_{SAW2}^0$. It shows that
there is a nonlinear quantum interaction between the SAW1 and
SAW2. This interaction is seen from the expression for $Q_1$

\begin{eqnarray}
\label{Qdot2}
Q_1=\frac{1}{\lambda^2}
\frac{1}{4\pi}N_t2\sum_{n,m,n_1,m_1}
\int dq_xdq_y <|U_0(q)|^2>_{Z_i}|A_{n,m,n_1,m_1}({\bf q})|^2
\\ \nonumber
c_sq_x(f^{0}_{n,m}-f^{0}_{n_1,m_1})
\delta(\hbar c_sq_x+\hbar c_sq_y+E_{n,m}-E_{n_1,m_1}).
\end{eqnarray}
In this equation the energies $E_{n,m}$ depend on $\Phi_{SAW1}^0$
and $\Phi_{SAW2}^0$. Physically, the first SAW interacts with the
second because the latter changes the energy spectrum in dots.
This interaction can be observed as giant quantum oscillations
\cite{PRL2001}.

Recent experiments \cite{Rocke}, which involved two SAW's
propagating in the perpendicular directions, were performed using
the pump-probe method. The first SAW had a high intensity and the
second was a probing wave. These experiments demonstrated strongly
nonlinear AE effects in the presence of photogenerated electrons
and holes.

\section{Intense acoustic waves in nanostructures}
\label{Dots2}

In this section, we briefly consider the nonlinear AE interaction
in a nanostructure which includes an array of quantum wires.
Electron quantum wires can be realized by etching a formerly
uniform 2D system \cite{wires}. In our model a SAW propagates
along the wire direction $x$ (Fig.6). In the $y$-direction, the
electron motion is finite due to the wire confinement. At high
intensity, the SAW can induce moving electron dots. The electron
potential near its minimum can be approximated by
$(m^*/2)(\Omega_{SAW}^2x'^2+\Omega_{wire}^2y^2)$, where
$\Omega_{wire}$ describes the static confinement in a quantum wire
in the $y$-direction and $\Omega_{SAW}$ is a function of the sound
intensity. The SAW absorption is given by Eq.~\ref{Qdot2} with the
corrections $\Omega_{SAW1}\rightarrow\Omega_{SAW}$,
$\Omega_{SAW2}\rightarrow\Omega_{wire}$ and $\delta(\hbar
c_sq_x+\hbar c_sq_y+E_{n,m}-E_{n_1,m_1})\rightarrow \delta(\hbar
c_sq_x+E_{n,m}-E_{n_1,m_1})$.

The calculated absorption of a SAW, $Q$, demonstrates giant
quantum oscillations due to the discrete spectrum in 'moving'
quantum dots (Fig.~6) . The minimums of $Q$ occur due to the
commensurability effect in the energy spectrum and correspond to
the condition $s\Omega_{SAW}=s'\Omega_{wire}$, where $s$ and $s'$
are integers. For example, we consider the main minimum in
$Q(I_{SAW})$ when $\Omega_{SAW}=\Omega_{wire}=1~meV$ ($s=s'=1$).
Under the condition $\Omega_{SAW}=\Omega_{wire}$, the spacing
between energy levels is equal to $\Omega_{SAW}=1~meV$, and the
absorption is dramatically suppressed.  This is due to the
exponential function (\ref{Qdot22}). When $\Omega_{SAW}$ slightly
differs from $\Omega_{wire}$, the minimum spacing between energy
levels becomes less and the absorption $Q$ increases.

Experiments with SAW's propagating along quantum wires have been
performed in Ref.~\cite{Santos}. In these experiments, SAW's drove
photogenerated electrons and holes through the quantum wires
fabricated on patterned interfaces.

\section{Discussion}
\label{C}

We see from our results that the acoustic absorption in an
electron system in the quantum limit is very different to that in
the classical case. In the classical plasma, $Q(\Phi_{SAW}^0)$
approaches $Q_{max}$ in the limit $\Phi_{SAW}^0\rightarrow\infty$
\cite{Rotter,Govorov00,3Dnonlin}. In the quantum case, the
absorption $Q(\Phi_{SAW}^0)$ first decreases with increasing
$\Phi_{SAW}^0$ and then becomes saturated at the minimal quantum
absorption which comes from residual scattering in the ground
subband. When two intense SAW's form electron quantum dots, the
absorption exponentially decreases. Thus, the dissipation of sound
in a system with moving dots vanishes in the limit of the high
acoustic intensity. This is due to the fully quantized spectrum in
the dots.

In first principle approach the electron conductivity
$\sigma_0=e\mu n_s$ can be calculated as an infinite series of the
"ladder" diagrams \cite{Abrikosov}. Another method to derive the
formula for $\sigma_0$ is based on the so-called Landauer dipole
\cite{Landauer}. In this method, the impurities induce the dipole
moments in the moving coordinate system in which the electron
plasma does not move. The resulting polarization should coincide
with the electric field in a sample.  The equality between the
electric field and the impurity-induced polarization permits to
find the conductivity.  Here we use another argument based on
dissipation. By using our arguments we can also obtain the formula
for the conductivity. To do this, we calculate the dissipation in
the moving coordinate system. Then, we compare the results for the
moving and static coordinates.

In conclusion, we have developed a quantum theory of the nonlinear interaction
between acoustic waves and a 2D electron system in the limit of the high
acoustic intensity.  The intense SAW creates electron quantum wires. The SAW
absorption as a function of the sound intensity reflects the 1D density of
states.
In wide wires the SAW absorption originates from the inter-subband transitions.
In the strictly 1D case there is still a residual SAW absorption, which comes
from the intra-subband electron transitions. In the regime of quantum dots the
SAW absorption approaches zero in the limit of high sound intensity. The
modern hybrid structures can be a candidate to experimentally observe the
described quantum nonlinear mechanisms of the acoustoelectric interaction.

We would like to thank A. Wixforth and J. P. Kotthaus for
important motivating remarks, and B. Laikhtman, A. V. Chaplik, A.
Mayer, and  H.-J. Kutschera for helpful discussions. We gratefully
acknowledge financial support by the Volkswagen-Foundation and by
the Russian Foundation for Basic Research.

$^*$ E-mail: Govor@isp.nsc.ru

$^{**}$ E-mail: Kalam@isp.nsc.ru

\newpage

{\bf Figure Captions}
\vskip 0.5 cm
{\bf Fig.~1} \\
{\bf a)} \ The cross section of a hybrid
semiconductor-piezoelectric structure. The electron density in a
quantum well depends on the transport gate voltage $V_t$. A SAW
propagates near the surface and interacts with the 2D plasma in a
quantum well.
\\ {\bf b)} \ The moving piezoelectric potential of a SAW and the energy structure of
moving quantum wires.
\\ {\bf c)} \ Intra-subband transitions in the 1D limit when electrons occupy
only the ground subband. The second diagram shows the inter-subband transitions
when electrons occupy the two lowest subbands.

\vskip 0.5 cm {\bf Fig.~2} \ Calculated SAW absorption as a
function of the potential amplitude $\Phi_{SAW}^0$ for various
temperatures for the case of $\delta$-impurities. $\lambda=1~\mu
m$, $N_L=4*10^{5}~cm^{-1}$, $N_s=4*10^9~cm^{-2}$, and
$N_t=4.1*10^{14}~cm^{-3}$. The low-temperature mobility in a
homogeneous 2D gas with $N_s=3*10^{11}~cm^{-2}$ is equal to
$3*10^6~cm^2/Vs$.

\vskip 0.5 cm {\bf Fig.~3} \ Calculated SAW absorption as a
function of the potential amplitude $\Phi_{SAW}^0$ for various
temperatures for the case of background Coulomb impurities which
are dynamically screened by electrons. The parameters are similar
to those of Fig.~2. Insert: the SAW absorption
$Q_{class}(\Phi_{SAW}^0)$ for the case of a classical electron
system in a quantum well at room temperature.

\vskip 0.5 cm
{\bf Fig.~4} \
The energy-loss rate due to acoustic phonon emission in quantum wires formed by
a SAW as a function of the electron temperature $T_e$ for various lattice
temperatures $T_{l}$ (solid curves). The dashed curves show the SAW dissipation
$Q(T_e)$ in the quantum regime for the case of Coulomb impurities. The dashed
curves 1 and 2 relate to $\mu_{2D}=1$ and $3*10^6~cm^2/Vs$, respectively.
The corresponding impurity densities are $N_t=12.3$ and $4.1*10^{14}~cm^{-3}$.
The 2D mobility is calculated for $N_s=3*10^{11}~cm^{-2}$. The crossing points
give the electron temperature in a wire. $\Phi_{SAW}^0=2~V$.

\vskip 0.5 cm
{\bf Fig.~5} \
The heating temperature $\Delta T=T_e-T_{l}$ in a wire as a function of the
mobility $\mu_{2D}$ for the system with Coulomb impurities; the lattice
temperature $T_l=2~K$; $\Phi_{SAW}^0=2~V$.
$\lambda=1~\mu m$ and  $N_L=4*10^{5}~cm^{-1}$.  The 2D mobility is calculated
for $N_s=3*10^{11}~cm^{-2}$.

\vskip 0.5 cm {\bf Fig.~6} \ Calculated SAW absorption as a
function of the quantization $\hbar\Omega_{SAW}^0$ in the system
with etched quantum wires; the energy $\hbar\Omega_{wire}$
describes the static quantization in the potential of wires.
Insert: Sketch of the system with etched quantum wires and a SAW.
The velocity of dots is $(c_s,0)$.

\newpage
\begin{figure}
\centering \psfig{file=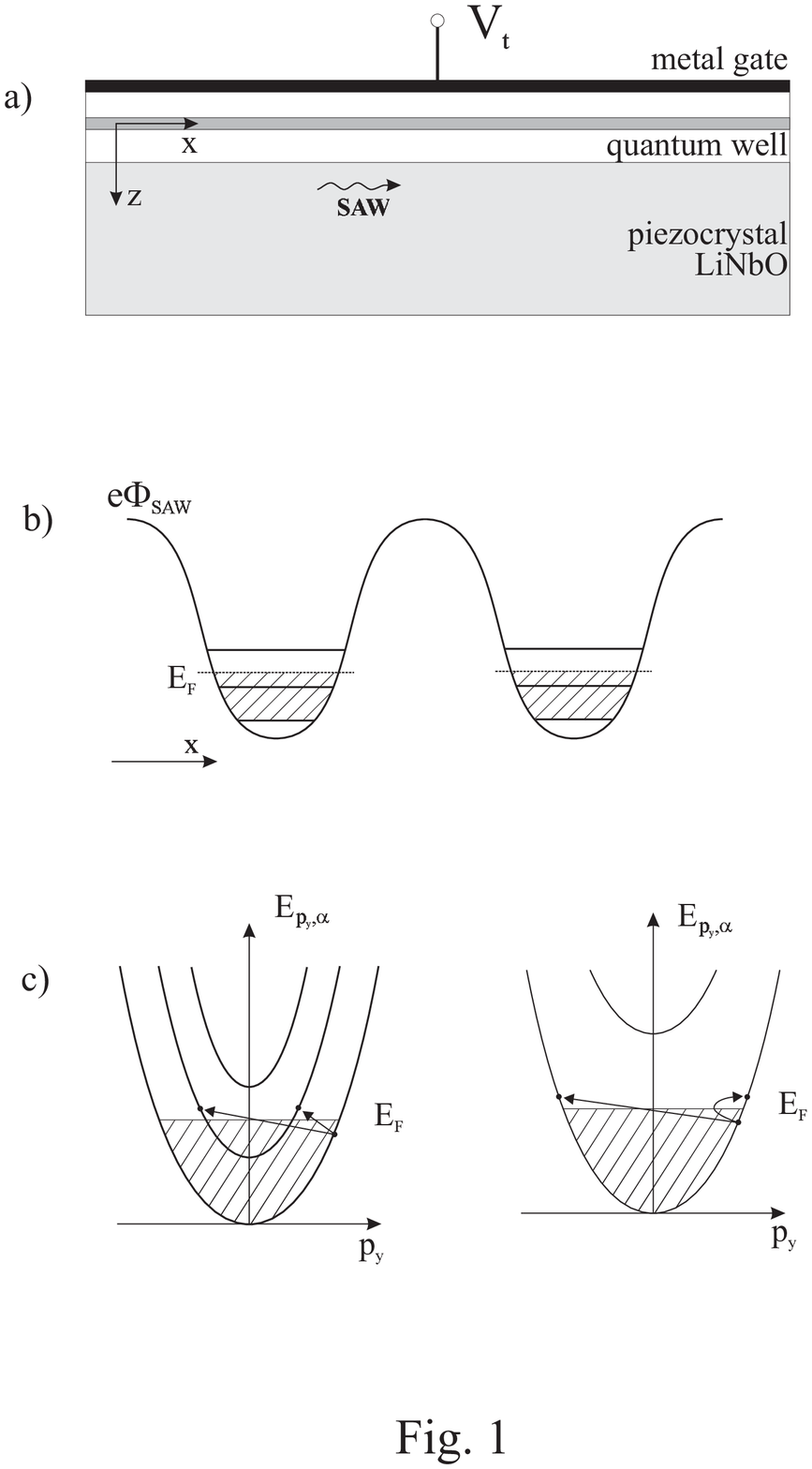,height=20cm}
\end{figure}

\newpage
\begin{figure}
\centering \psfig{file=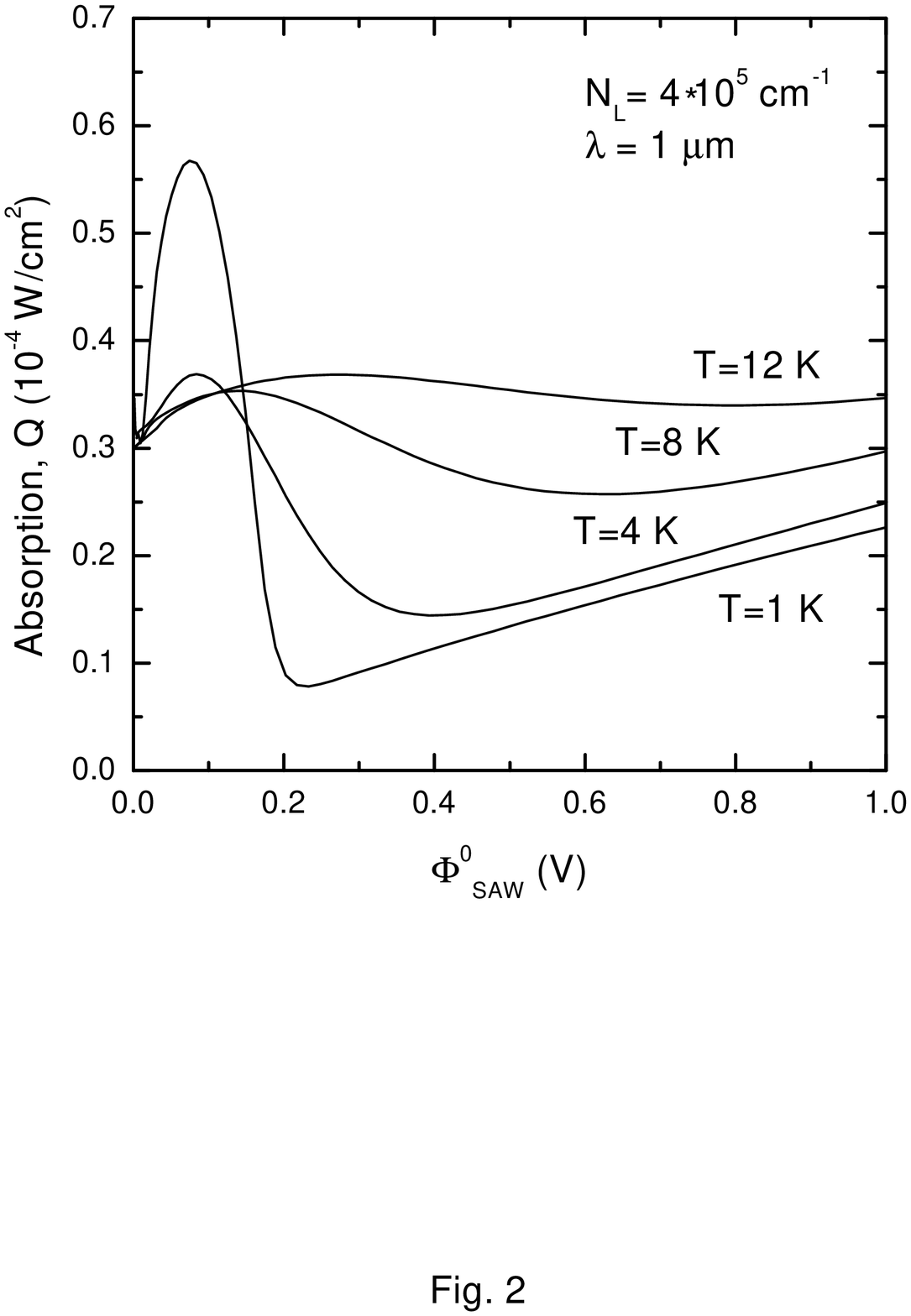,height=20cm}
\end{figure}

\newpage
\begin{figure}
\centering \psfig{file=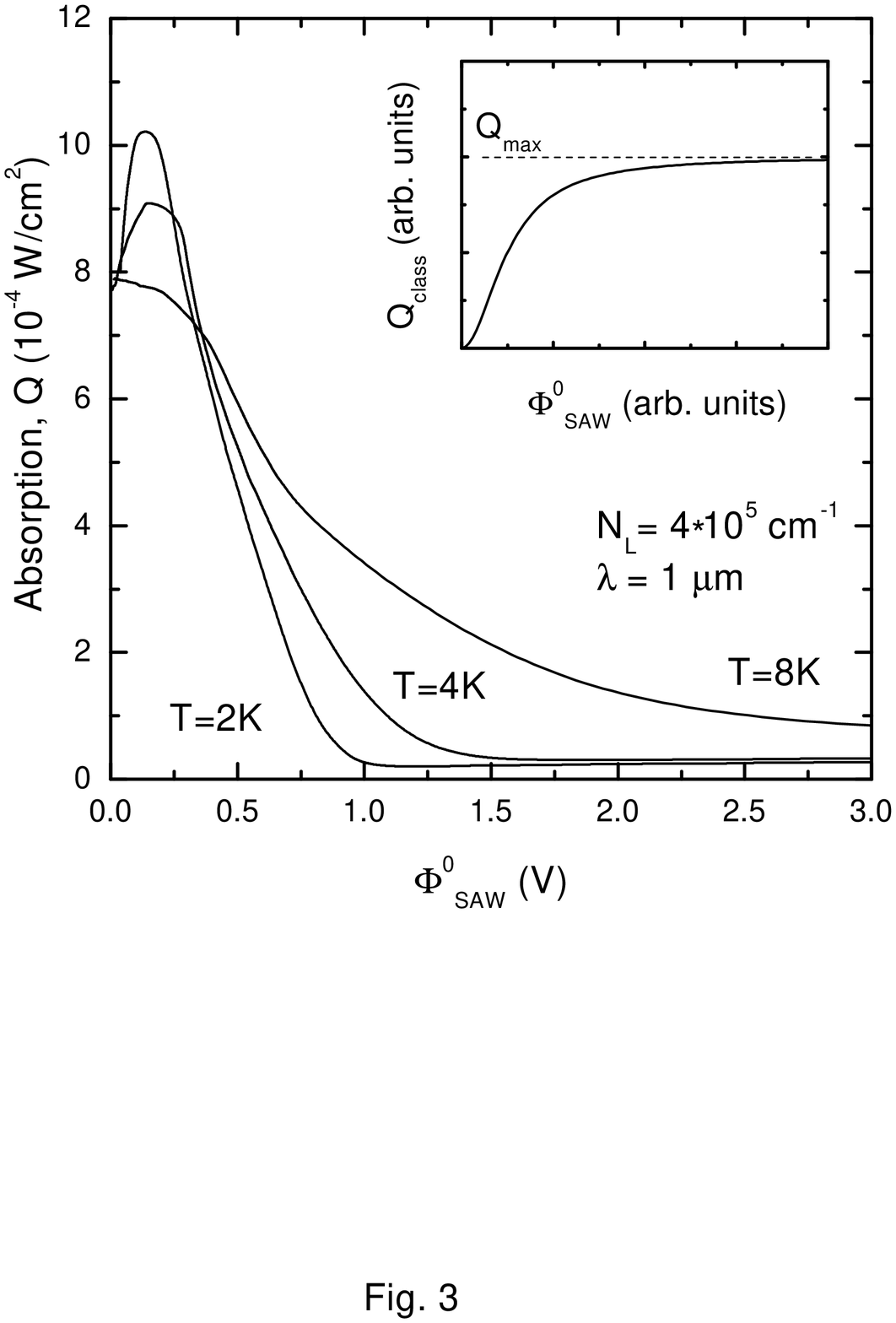,height=20cm}
\end{figure}

\newpage
\begin{figure}
\centering \psfig{file=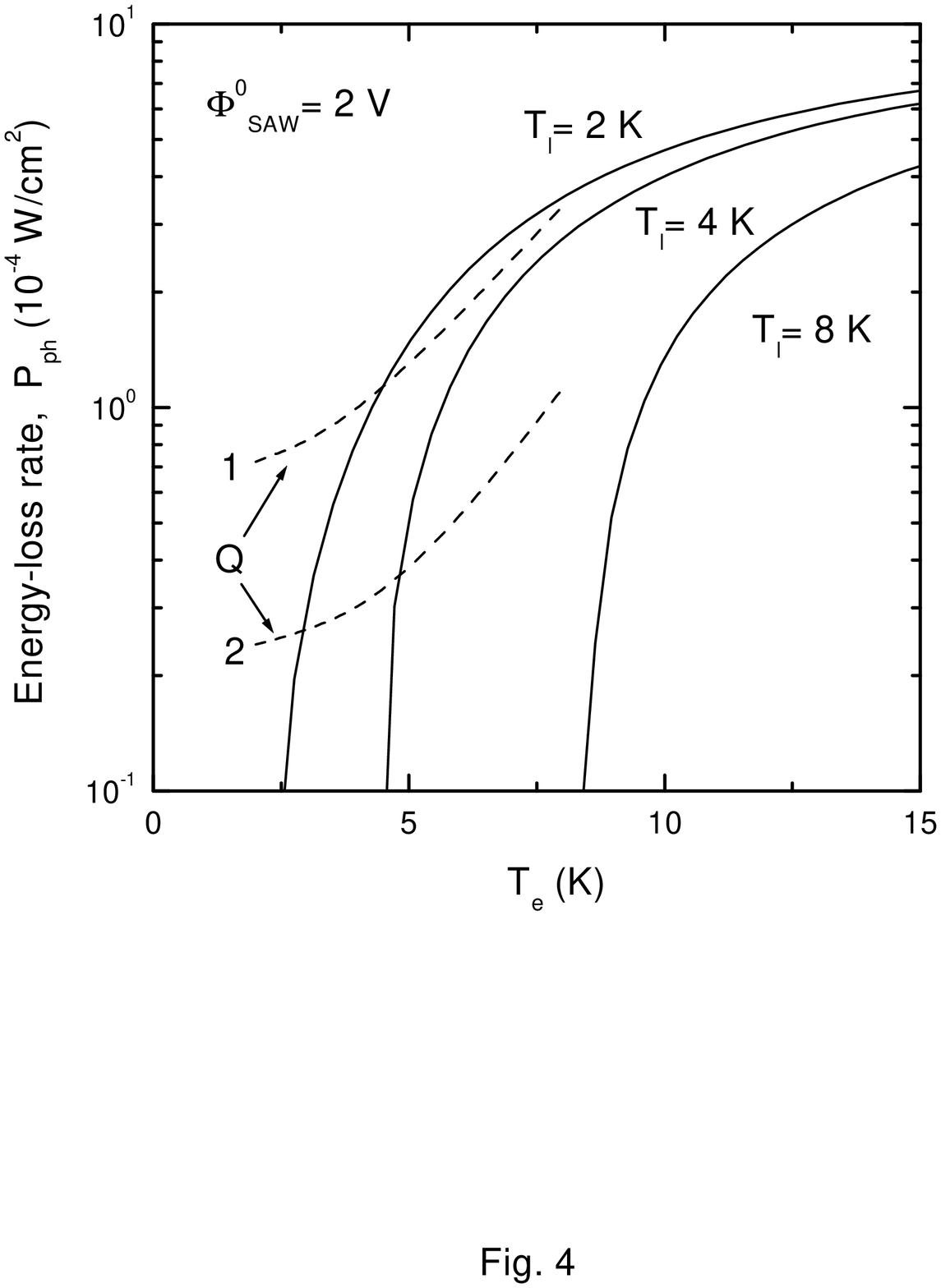,height=20cm}
\end{figure}

\newpage
\begin{figure}
\centering \psfig{file=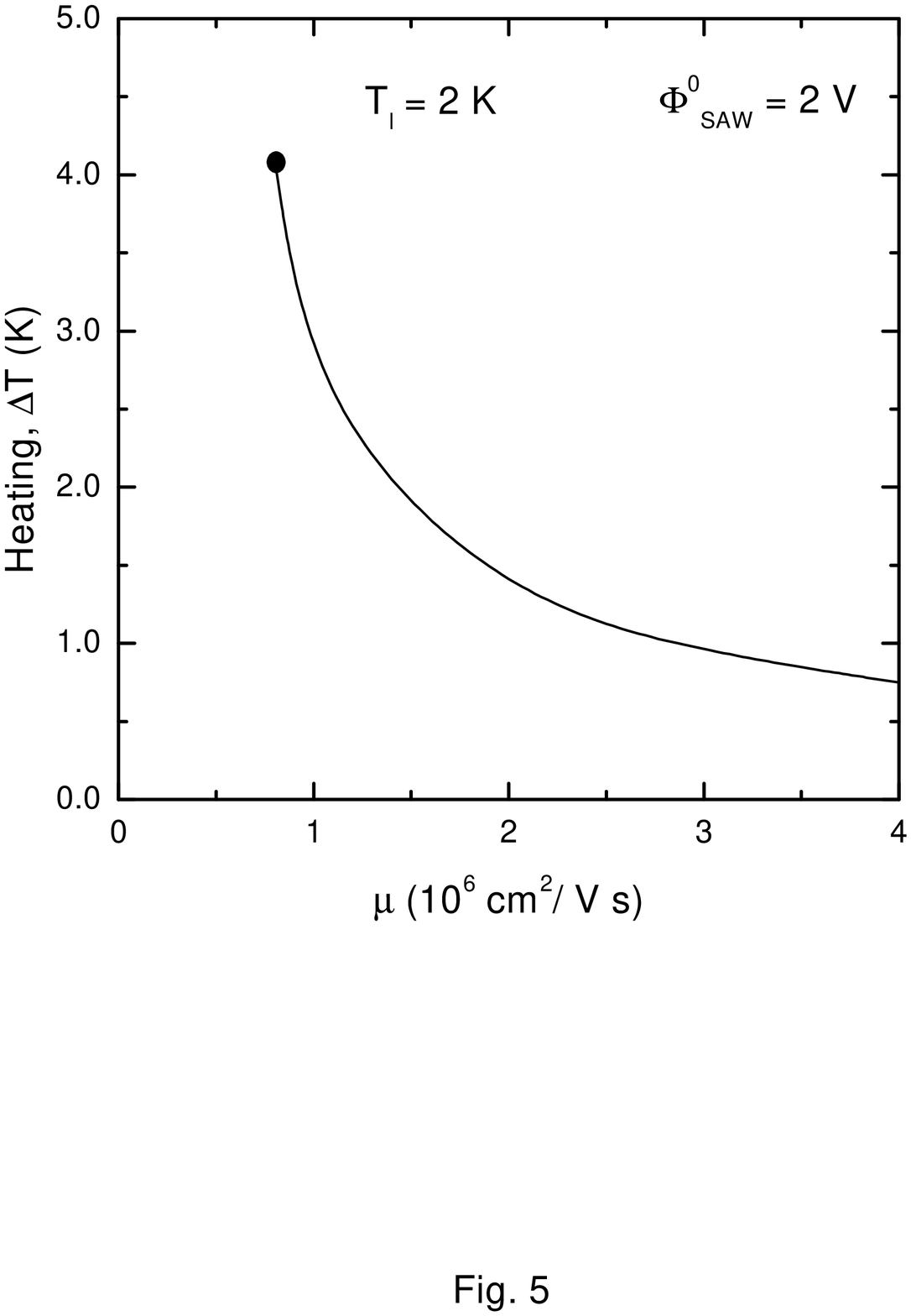,height=20cm}
\end{figure}

\newpage
\begin{figure}
\centering \psfig{file=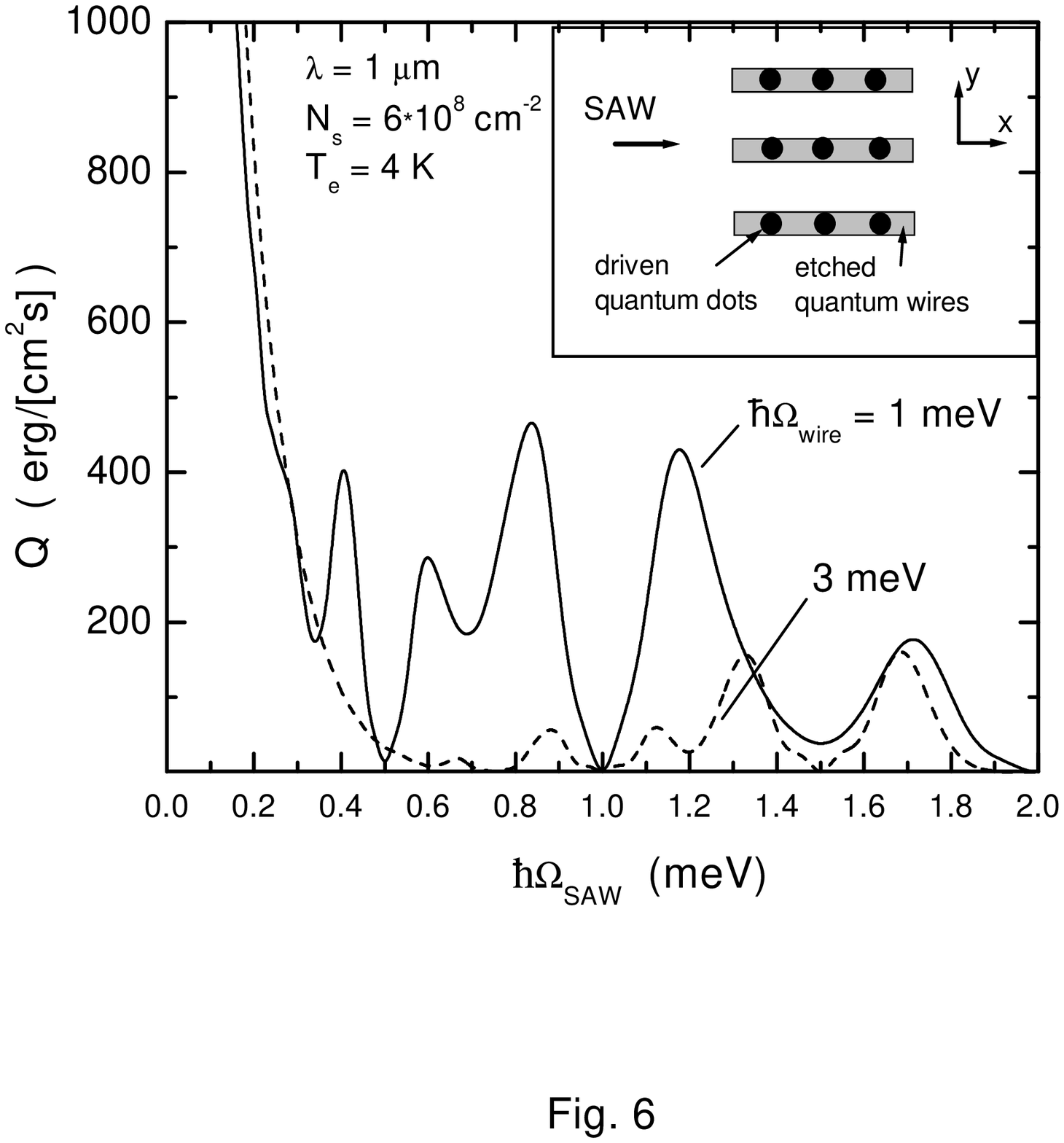,height=20cm}
\end{figure}


\begin{thebibliography}{31}

\bibitem{Wees-WharamJPCM}
B. J. van Wees et al., Phys. Rev. Lett. {\bf 60}, 848 (1988);
D. A. Wharam et al., J. Phys. {\bf C 21}, L209 (1988).

\bibitem{Landauer-Buttiker}
R.~Landauer, IBM J. Res. Dev. {\bf 1}, 223 (1957); R.~Landauer,
Phys. Lett. {\bf 85A}, 91 (1981); B.I.~Halperin, Phys. Rev. B {\bf
25}, 2185 (1982); M.~B\"uttiker, Phys. Rev. Lett. {\bf 57}, 1761
(1986).

\bibitem{thermal}
L.G.C.~Rego and G.~Kirczenow, Phys. Rev. Lett. {\bf 81}, 232
(1998); K.~Schwab, E.A.~Henriksen, J.M.~Worlock, and M.L.~Roukes,
Nature, {\bf 404}, 974 (2000).

\bibitem{theory}
J.~M.~Luttinger, J. Math. Phys. {\bf 4}, 1154 (1963);
J.~Voit, Rep. Prog. Phys. 58, 977 (1995);
L.~Calmels and A.~Gold, Phys. Rev. B 56, 1762 (1997).

\bibitem{Govorov91}
A.O.~Govorov and A.V.~Chaplik, Zh. Eksp. Teor. Fiz. {\bf 99}, 1853
(1991) [Sov. Phys. JETP {\bf 72}, 1037-1046 (1991)].


\bibitem{Hawrylak}
L.~Jacak, P.~Hawrylak, and A.~Wojs, {\it Quantum dots}
(Springer, Berlin, 1998);
M. Grundmann, D. Bimberg, and  N. N. Ledentsov,
{\em Quantum Dot Heterostructures} (Wiley, New York, 1998).

\bibitem{wires}
W. Hansen, M. Horst, J. P. Kotthaus, U. Merkt, Ch. Sikorski,
and K. Ploog, Phys. Rev. Lett. {\bf 58}, 2586 (1987);
E. Batke, D. Heitmann, and C. W. Tu, Phys. Rev. B {\bf 34}, 6951 (1986);
T. P. Smith III et al., Phys. Rev. Lett. {\bf 59}, 2802 (1987);
T. Egeler, G. Abstreiter, G. Weimann , T. Demel, D. Heitmann, P. Grambow , and
W. Schlapp,
Phys. Rev. Lett. {\bf 65}, 1804 (1987);
A. R. Goni, A. Pinczuk, J. S. Weiner,
B. S. Dennis, L. N. Pfeiffer, and K. W. West, Phys. Rev. Lett. {\bf 70}, 1151
(1993);
H. Drexler et al., Phys. Rev. B {\bf 49}, 14074 (1994).


\bibitem{Rotter}
M.~Rotter, A.~V.~Kalameitsev, A.~O.~Govorov, W.~Ruile,
and A.~Wixforth, Phys. Rev. Lett. {\bf 82}, 2171 (1999);
M.~Rotter, A.~Wixforth, A.~O.~Govorov, W.~Ruile,
D.~Bernklau, and H.~Riechert,  Appl. Phys. Lett. {\bf 75}, 965 (1999).

\bibitem{Govorov00}
A.O.~Govorov et al., Phys. Rev. {\bf B 62}, 2659 (2000).

\bibitem{3Dnonlin}
P.K.~Tien, Phys. Rev. {\bf 171}, 970 (1968);
V.L.~Gurevich and B.D.~Laikhtman,
Zh. Eksp. Teor. Fiz. {\bf 46}, 598 (1964)
[Sov. Phys. JETP {\bf 19}, 407 (1964) ];
Yu.V.~Gulyaev,
Fiz. Tverd. Tela {\bf 12}, 415 (1970)
[Sov. Phys. - Solid State,
{\bf 12}, 328 (1970)].


\bibitem{Keldysh} L.V.~Keldysh, Fiz. Tverd. Tela {\bf 4}, 1015 (1962)
[Sov. Phys. - Solid State].

\bibitem{Popov}
V.V.~Popov and A.V.~Chaplik,
Zh. Eksp. Teor. Fiz. {\bf 73}, 1009 (1977)
[Sov. Phys. JETP].

\bibitem{Laikhtman}
B.D.~Laikhtman and Yu.V.~Pogorel'skii,
Zh. Eksp. Teor. Fiz. {\bf 75}, 1892 (1978)
[Sov. Phys. JETP, 8, 953 (1978)].

\bibitem{Wixforth89}
A.~Wixforth et al., Phys. Rev. {\bf B40}, 7874 (1989);
R.L.~Willett et al., Phys. Rev. Lett. {\bf 71}, 3846 (1993);
I.L.~Drichko et al., Fiz. Tekh. Poluprovodn. {\bf 31}, 451 (1997)
[Sov. Phys. Semiconductors {\bf 31}, 451 (1997)].

\bibitem{Talyanskii}
V.I.~Talyanskii et al., Phys. Rev. {\bf B56}, 15180 (1997);
J.M.~Shilton et al., Phys. Rev. {\bf B51}, 14770 (1995).

\bibitem{2Dlin}
K.A.~Ingebrigtsen, J. Appl. Phys. {\bf 41}, 454 (1970);
A.V.~Chaplik, Pis. Zh. Tekh. Fiz. {\bf 10}, 1385 (1984)
[Sov. Tech. Phys. Lett. {\bf 10}, 584 (1984)].

\bibitem{nanolin}
V.L.~Gurevich, V.B.~Pevzner, and G.J.~Iafrate, Phys. Rev. Lett. {\bf 77},
3881 (1996);
A.D.~Mirlin and P.~W\"olfle,  {\it ibid} {\bf 78}, 3717 (1997);
Y.~Levinson et al.,  Phys. Rev. {\bf B 58}, 7113 (1998);
C.~Eckl, Yu.A.~Kosevich, and A.P.~Mayer, Phys. Rev. {\bf B 61}, XXX (2000).


\bibitem{Aizin} G.R.~Aizin, G.~Gumbs, and M.~Pepper,
Phys. Rev. {\bf B 58}, 10 589 (1998).


\bibitem{QuinnJJ}
M.P.~Greene et al., Phys. Rev. {\bf 177}, 1019 (1969);
K.S.~Yu and J.J.~Quinn, {\it ibid}, {\bf B 27}, 1184 (1983).

\bibitem{Weinreich}
G.~Weinreich,  Phys. Rev. {\bf 107}, 317 (1957). ~

\bibitem{Qmax}
The formula for $Q_{max}$ can easily be obtained using the hydrodynamic
equations. Within the Drude model, the friction force acting on an electron is
equal to $f_{fr}=-v(r,t)m^*/\tau_{tr}$, where $v(r,t)$ is the local velocity in
a 2D plasma.
The SAW dissipation is $Q_{max}=-<f_{fr}v n_s>$. It follows from the
conservation of charge that in the regime of separated wires $v(x,t)=c_s$ and so
$Q_{max}=m^*c_s^2*N_s/\tau_{tr}$.

\bibitem{SNOSKA}
At low temperatures electrons driven by a SAW can not emit bulk acoustic phonons
because $c_s<c_{l,t}$, where $c_{l,t}$ are the bulk phonon velocities. Thus, the
phonon contribution to $Q$ can be neglected.

\bibitem{Davies}
J.H.~Davies, {\it The Physics of Low-Dimensional Semiconductors}
(Cambridge, University-Press, UK, 1998).


\bibitem{Abrikosov}
A.A.~Abrikosov, L.P.~Gorkov, and I.E.~Dzyaloshinskii,
{\it Methods of Quantum Field Theory in Statistical Physics}
(Prentice-Hall, Englewood Cliffs, NJ, 1963).

\bibitem{Bockelmann}
U.~Bockelmann and G.~Bastard, Phys. Rev. {\bf B 42},  8947 (1990);
C.~Kurdak et al., Appl. Phys. Lett. {\bf 67},  386 (1995).

\bibitem{potassium}
M.~Zgonik et al., J. Appl. Phys. {\bf 74}, 1287 (1993).

 \bibitem{PRL2001}
A.O.~Govorov, A.V.~Kalameitsev, V.M.~Kovalev, H.-J.~Kutschera, and
A.~Wixforth, Phys. Rev. Lett. 87, 226803 (2001).

\bibitem{Rocke}
C.~Rocke, A.O.~Govorov, A.~Wixforth, G.~B\"ohm, and G.~Weimann,
Phys. Rev. {\bf B 57},  R6850 (1998);
M.~Streibl, A.~Wixforth, J.P.~Kotthaus, A.O.~Govorov, C.~Kadow, and
A.C.~Gossard, Appl. Phys. Lett. 75, 4139 (1999).

\bibitem{Santos}
P.~Santos et al., Proceedings of ICPS-2002, Edinburgh.

\bibitem{Landauer}
R.~Landauer, Z. Phys. {\bf B 21},  247 (1975); W.~Zwerger et al.,
Phys. Rev. {\bf B 43},  6434 (1991).


\end{thebibliography}
\end{document}